\documentclass[12pt]{article}

\def\ben{\begin{enumerate}}  \def\een{\end{enumerate}}
\def\beq{\begin{equation}}   \def\eeq{\end{equation}}
\def\bea{\begin{eqnarray}}  \def\eea{\end{eqnarray}}
\def\nn{\nonumber}
\def\noi{\noindent}

\def\lsim{\raise0.3ex\hbox{$<$\kern-0.75em\raise-1.1ex\hbox{$\sim$}}}
\def\gsim{\raise0.3ex\hbox{$>$\kern-0.75em\raise-1.1ex\hbox{$\sim$}}}

\begin{document}
\begin{flushright}
CERN-TH/2002-128
\end{flushright}
\begin{center}
{\bf BOUND STATES IN ONE AND TWO SPATIAL DIMENSIONS}
\vspace{0.25 truecm}

{\bf K. Chadan}\\
{\it Laboratoire de Physique Th\'eorique}\footnote{Unit\'e Mixte de Recherche
UMR 8627 - CNRS }\\    {\it Universit\'e de Paris XI, B\^atiment 210, 91405
Orsay Cedex, France} \par \vskip 2 truemm

{\bf N. N. Khuri} \\
{\it Department of Physics, The Rockefeller University, New York, NY
10021, U.S.A.} \par \vskip 2 truemm

{\bf A. Martin}\\
{\it TH Division, CERN, Geneva, Switzerland and}\\
{\it Laboratoire de Physique Th\'eorique, F-74941 Annecy-le-Vieux,
France} \par \vskip 2 truemm

{\bf Tai Tsun Wu} \\
{\it Gordon McKay Laboratory, Harvard University, Cambridge, MA 02138,  
U.S.A.}\\
{\it and TH Division, CERN, Geneva, Switzerland}
   \end{center}
\vskip 0.25 truecm

\begin{abstract}
In this paper we study the number of bound states for potentials
in one and two spatial dimensions. We first show that in addition
to the well-known fact that an arbitrarily weak attractive potential
has a bound state, it is easy to construct examples where weak
potentials have an infinite number of bound states. These examples
have potentials which decrease at infinity faster than expected.
Using somewhat stronger conditions, we derive explicit bounds on
the number of bound states in one dimension, using known results
for the three-dimensional zero angular momentum. A change of
variables which allows us to go from the one-dimensional case to
that of two dimensions results in a bound for the zero angular
momentum case. Finally, we obtain a bound on the total number of
bound states in two dimensions, first for the radial case and then,
under stronger conditions, for the non-central case.

   \end{abstract}

\begin{flushleft}
LPT Orsay 02-57 \\
CERN-TH/2002-128\\
June 2002
\end{flushleft}

\newpage
\pagestyle{plain}

\noi {\large\bf I. Introduction}\\

In recent years, it has become apparent that studying physics in two spatial
dimensions is not just an academic exercise, especially for condensed
matter physics where there are bound states
due to impurities on the surface of a semiconductor or at a junction
\cite{1r}. In addition, we have established a
remarkable universality property for low energy scattering in two
dimensions. Namely, excluding some well-defined
and rare exceptional cases, the $m = 0$ phase shift for a radial
potential behaves like $(\pi /2)  (\ell nk)^{-1}$
as $k \to 0$ \cite{2r}. This result has been recently generalized to
non-radial and even non-local potentials
\cite{3r}.  \par

We believe that relatively little is known about bound states in one
and two dimensions. For any
dimension, including one and two, we know that if the potential is
sufficiently smooth and
sufficiently rapidly decreasing at large distances, there is a
semi-classical asymptotic estimate of
the number of bound states for a potential $gV,g \to \infty$, which
was first established for the radial case in
\cite{4r}, then generalized in \cite{5r} to arbitrary
dimensions.\par

However, concerning strict bounds on the number of bound states the
situation is radically different
for one and two dimensions from that in higher dimensions (including
three dimensions). Lieb \cite{6r},
Cwikel \cite{7r} and Rozenblum \cite {8r} have shown that for $n \geq
3$, $n$ being the number of spatial
dimensions, there is a bound

\beq
\label{1e}
N \leq B_n \int |V|^{n/2}\ d^n x \ ,
  \eeq

\noi where $B_n$ is definitely {\it larger}, even for very large
dimensions, contrary to earlier belief \cite{9r}, than the semi-classical
constant $C_n$ appearing in the asymptotic estimate \cite{5r}

\beq
\label{2e}
N(g) \sim C_n \ g^{n/2} \int (V^-)^{n/2} \ d^n x \quad , \quad g \to  
\infty \ ,
C_n = \frac{2^{-n} \pi^{-n/2}}{\Gamma (1 + \frac{n}{2})}
\eeq

\noi for a potential $gV$ where $- V^-$ is the negative part of the
potential~: $V = V^+ - V^-$, $V^{\pm} \geq
0$.   For central potentials, $B_n / C_n \to 1$ for $n \to \infty$  
\cite{9r}. Other proofs have been
obtained
\cite{9r},
\cite{10r}.  Furthermore, it is well known that for one and two
dimensions a potential globally attractive, arbitrarily weak, such
that

\beq
\label{3e}
\int d^n x V(x) < 0 \ , \ n = 1,2
		 \eeq

\noi has a bound state. The proof is trivial for $n = 1$ by using a
Gaussian trial function. For $n=2$,
there is a proof by Simon, for instance \cite{11r}. The simplest one is by 
Yang and
De Llano \cite{12r} who use a trial
function $\exp - (r + r_0)^{\alpha}$, $\alpha$ sufficiently small. \par

However, this bound state has an incredibly small binding energy in
absolute value, for a potential
$gV$, which behaves like $\exp - {c \over g}$ for small $g$, as shown
in Appendix I. \par

In addition to the above, we note that for the $s$-state ($m = 0$),
and $n = 2$, there is an old bound on the
number of bound states due to Newton \cite{13r} and Set\^o
\cite{14r}. However, this bound is bilinear in $V$ and
does not behave like the semiclassical result for large $g$. \par

It was noticed in Ref. \cite{9r} that the number of bound states in
two dimensions is certainly larger
than $-{1 \over 4} \int rV(r)dr$, in the central case. \par

In this paper we first find examples of potentials in one dimension
for which the number of bound
states is infinite. Using a transformation which is systematically
studied, one can find more refined
potentials for which the number of bound states is infinite. \par

This same transformation allows us also to find radial potentials in
two dimensions for which the zero
angular momentum bound states are infinite in number. Examples with
non-radial potentials are also
constructed. All these examples possess the property, $\int d^2x
|V(\vec{x})| < \infty$, and in addition
$\int d^2x |V(\vec{x}) | \ell n (2 + |\vec{x}|)^{1- \varepsilon} <
\infty$.\par

In section III we find explicit bounds on the number of bound states
in one dimension by using
well-known bounds for the three dimensional radial case with zero
angular momentum. In addition, using
the above noted change of variables, we also obtain bounds on the
number of zero angular momentum bound
states in two dimensions. \par

Finally, in section IV, we get bounds on the {\it total} number of
bound states in two dimensions. This
bound has the property that it is linear in $g$ for a potential $gV$
and is thus similar to the
semi-classical estimate. \par

In Appendix I we give upper and lower bounds on the ground state
energy in two dimensions. \par

Next, in Appendix II, we present a system of transformations which
first allow us to derive more and
more refined examples of limit potentials with a finite or infinite
number of bound states. Secondly,
these transformations allow us to convert results obtained in a given
dimension to results for another
dimension for zero angular momentum. \par

In Appendix III we compare one of our two dimensional bounds with the
Newton-Set\^o bound. Finally, in
Appendix IV, we sketch the proof that bound states are on real
analytic Regge trajectories. \cite{15r} \par

A preliminary account of these results was presented at a workshop
in Les Houches \cite{16r}.\\

\noi {\large\bf II. Examples where the number of bound states is infinite}\\

We begin by using the well known result that in one dimension, and
for the radial case in 2 and 3 dimensions, the
number of negative energy bound states is equal to the number of
nodes of the zero energy wave-function
\cite{17r}. \par

For any two
potentials $V_1(x) \leq 0$, and $V_2
(x) \leq 0$ in one dimension, one can easily show that, if $V_1(x) >  
V_2(x)$, then for
any interval $a \leq x \leq b$, we have

\beq
\label{4e}
n_2(a,b) \geq n_1(a,b) - 1 \ ;
		 \eeq

\noi where $n(a,b)$ is the number of nodes in the interval $(a,b)$.
Thus if $n_1(x, \infty )$ is infinite,
$n_2(x, \infty)$ is also infinite. \par

We write the zero energy one-dimensional Schr\"odinger equation for
an attractive potential $V =
-\lambda /x^2$~; $x > x_0 > 0$, $\lambda >0$,

\beq
\label{5e}
\left ( - {d^2 \over dx^2} - {\lambda \over x^2} \right ) \phi (x) = 0 \ .
		 \eeq

\noi Because of the homogeneity of Eq. (\ref{5e}), $\phi = x^s$,
where $s$ is given by the two roots
$s_{\pm}$ of the equation
$$
s(s - 1) = - \lambda ,
$$
{\rm or}
\beq
\label{6e}
s_{\pm} = {1 \over 2} \pm \sqrt{{1 \over 4} - \lambda} \ .
\eeq

\noi For $\lambda > 1/4$, both $s_+$ and $s_-$ are complex, and the
solution $\phi$ can be constructed by taking a
linear combination of $x^{s_+}$ and $x^{s_-}$. We have

\beq
\label{7e}
\phi (x) = \sqrt{x} \cos \left ( \sqrt{\lambda - 1/4} \log x +
\delta \right ) \ .
		 \eeq

\noi Obviously, this $\phi$  has an infinite number of nodes for any $X 
\leq x < \infty$, $X > 0$.\par

We can now use the theorem summarized in Eq. (\ref{4e}), to get the
following general result~: the number
of one dimensional bound states is infinite if there exists an $X >
0$ such that
\bea
			\label{8e}
&\hbox{either} \quad &x^2 V(x) < L < - 1/4\ , \ \hbox{for}\ x > X \ , \nn \\
&\hbox{and/or} \quad &x^2V(x) < L < - 1/4 \ , \ \hbox{for}\ x < -X \ .
\eea

\noi On the other hand, if $V$ is \underbar{bounded from below} and
if $x^2 V(x) > - 1/4$ for $|x| > |X|$, then
the number of bound states is finite. \par

Using the series of transformation described in Appendix II it is
possible to approach the limiting case
in a more refined way. For example~: if
$$
V(x) < - {1 \over 4x^2} - {\mu_1 \over 4x^2(\ell n x)^2}\ ,
\ x > X, \mu_1 > 1
$$
{\rm or} \\
\beq
\label{9e}
V(x) < - {1 \over 4x^2} - {1 \over 4x^2(\ell
n x)^2} \left [ 1 + {\mu_2 \over (\ell n
\ell n x)^2} \right ] \ , \ x > X, \mu_2 > 1
\eeq

\noi the number of bound states is {\it infinite}. Notice that this
is true for $X$ arbitrarily large, i.e.,
in a way, $V$ \underline{arbitrarily small}.\par

These two examples are such that $\int dx |V(x)|^{1/2} \to \infty$.
This is not surprising since in the
three dimensional radial case we have for a monotonic potential the
Cohn-Calogero \cite{18r} bound,
  \beq
\label{10e}
n < {2 \over \pi} \int_0^{\infty} d r |V|^{1/2} \ .
  \eeq

However we can have non-monotonic potentials such that the above
integral converges but the number of
bound states is infinite. For example one can set

\beq
\label{11e}
V = -  \sum_{0}^{+\infty} \delta (x - 2n) \ .
			\eeq

\noi For this potential $\int |V|^{1/2}dx = 0$ since the
$\delta$-function can be effectively replaced by
suitably chosen square wells of decreasing widths $\varepsilon_n$ and
depth ${1 \over \varepsilon_n}$ with $\Sigma
\sqrt{\varepsilon_n}$ convergent, and $\varepsilon_o$ arbitrarily small. 
\par

Next we consider the two-dimensional case. In this case we introduce
a simple transformation which converts
the {\it one}-dimensional zero energy Schr\"odinger equation to the
$m = 0$, {\it two}-dimensional radial
Schr\"odinger equation. In one dimension $- \infty < x < + \infty$ we have,

\beq
			\label{12e}
\left [ - {d^2 \over dx^2} + U(x) \right ] \phi (x) = 0 \ .
\eeq

\noi Our change of variables is given by~:

\bea
			\label{13e}
&&x \equiv \ell n \ r/R \  , \quad 0 \leq r < \infty \ ; \nn \\
&&U(x) \equiv r^2 \ V(r) \ ; \nn \\
&&\phi (x) = \psi (r) \ .
\eea

\noi This transformation is a particular case of the Liouville
transformation \cite{19r}. Equation (\ref{12e}) now
becomes

\beq
			\label{14e}
\left ( - {d^2 \over dr^2} - {1 \over r}\ {d \over dr}  + V(r) \right
) \phi (r) = 0 \ .
\eeq

\noi But this equation is precisely the $m = 0$ {\it two}-
dimensional radial Schr\"odinger equation. \par

Using Eq. (\ref{8e}) we now see that for a radial potential $V(r)$,
the number of bound states is
infinite if,

\beq
			\label{15e}
r^2 \left (\ell n {r\over R}\right )^2 V(r) < L < - 1/4 \ ; \quad r >
R_0 > R \ ;
  \eeq

This time we see that the integral appearing in the semiclassical
estimate, $\int_0^{\infty} r |V(r)|dr$
is convergent and yet the number of bound states is infinite.
Furthermore the integral $\int_0^{\infty}
r dr |V(r)|[\ell n (2 + r)]^{1 - \varepsilon }$, is also convergent
for $\varepsilon > 0$, and the integral can
be made arbitrarily small by taking $R_0$ arbitrarily large. \par

Our limit potentials in the 2-dimensional case are given by

\begin{eqnarray}
V(r) &=& - {\mu \over 4} \ {1 \over r^2\left (\ell n {r\over R}\right
)^2} \ , \quad r \geq R_0 \geq 1\ ; \nonumber \\
V(r) &=& 0  \hskip 1 truecm , \quad r < R_0 \ ,
\end{eqnarray}

\noi with $\mu > 1$.

In addition we can also solve the Schr\"odinger
equation exactly for the class,

\beq
			\label{16e}
V(r) = \left \{ \begin{array}{l} 0 \quad , \quad r < R\ , {\rm with} \  
R > 1 \ ;
\\ - g /r^2(\ell n
r)^{\alpha} \ , \ r > R\ , \ 1 < \alpha < 2  \end{array}\right .  \eeq

\noi with $g > 0$. The solution is given by

\beq
			\label{17e}
\left \{ \begin{array}{l}
\psi (r) = a + b \ \ell n r \ ; \quad r < R \ ; \\ \\
  \psi (r) = (\log r )^{1/2} \left [ A J_{\nu} \left ( 2 \nu \sqrt{g}
(\log r)^{1/2\nu} \right
) + B Y_{\nu}  \left ( 2\nu \sqrt{g} (\log r )^{{1 \over 2 \nu}} \right ) 
\right ] \ , \ r \geq R\
.\end{array} \right .\eeq

\noi where $\nu \equiv (2 - \alpha )^{-1}$, and $J_{\nu}$ and
$Y_{\nu}$ are Bessel functions. This
last solution has an infinite number of nodes for $r > R$ and hence
the potential (\ref{16e}) has an infinite
number of bound states, and this is true for arbitrarily small $g$. \par

A completely different approach to get infinitely many bound states
abandons radial symmetry and
considers scattering by circular ``delta shell'' potentials in the
plane. Indeed a very simple example
where $\int V d^2x$ is finite, arbitrarily small, and where one sees
that has a bound state has been
invented by Richard \cite{20r}. It is a delta shell potential~:

\beq
			\label{18e}
V = - g \delta (r - 1) \ .
\eeq

\noi Here $\int d^2xV = - 2 \pi g$ is finite. The zero-energy
Schr\"odinger equation

$$\left ( - {d^2 \over dr^2} - {1 \over r} \ {d \over dr} + V \right
) \psi = 0$$

\noi has a solution, finite at the origin, which is

  \beq
			\label{19e}
\left \{ \begin{array}{l} \psi = 1 \quad \hbox{for}\ r < 1 \\ \psi =
1 - g \ \ell n\  r \quad \hbox{for} \ r
\geq 1 \end{array} \right . \ .\eeq

\noi Hence the zero-energy radial solution has a node at

\beq
			\label{20e}
r_0 = \exp {1 \over g} \ ,
\eeq

\noi and therefore this potential has a bound state for arbitrarily
small $g$. \par

If, in addition, we now impose a Dirichlet boundary condition at $r =
\exp {1 \over g}$ and set $\psi$ to
be identically zero for $r > \exp \left (  {1 \over g}\right )$,
i.e., physically, having an infinitely
repulsive wall, we will still have a solution with a node at $r =
\exp {1 \over g}$, and hence a
zero-energy bound state.\par

Take now a sequence of potentials

\beq
			\label{21e}
V_n = - g_n \ \delta (|\vec{x} - \vec{x}_n|-1),
\eeq
\noi $g_n >0$,
such that $\Sigma \ g_n$ converges, $\vec{x}_n$ on the positive
$x$ axis. For simplicity, $g_n$ will
be chosen a decreasing sequence.
It is always possible to choose the $\vec{x}_n$'s in such a way that  
the disks

\beq
			\label{22e}
|\vec{x} - \vec{x}_n| \leq \exp {1 \over g_n} = r_n
\eeq
\noi do not overlap. \par

The number of bound states of $V = \sum\limits_{n=0}^{n_0} V_n$ is
certainly larger than $n_0$, the
result one gets when one imposes Dirichlet boundary conditions on the
border of each disk (this
strategy was already used in Ref. \cite{5r}). Letting $n_0$ go to
infinity, we see that we have
infinitely many bound states, and yet the integral $\int |V| d^2x = 2
\pi \Sigma g_n$ is finite and can
be arbitrarily small. \par

We can, however, do better than that, i.e., try to build an example in which

$$\int |V|\left [ \ell n (2 + |\vec{x}|)\right ]^{\alpha} \ d^2x$$

\noi is finite, where $\alpha$ is to be determined. We take the centres 
of the circles  on a line, and
since the $g_n$'s are decreasing, we have

$$|\vec{x}_n| + r_n < (2n + 1) \exp {1 \over g_n} \ ,$$

\noi and hence
$$\int|V_n| \ \left | \ell n (2 + |\vec{x}|)\right |^{\alpha}	\
d^2x < g_n \ \ell n \left [ 2 + (2n + 1) \exp
{1 \over g_n}\right ]^{\alpha} \ .$$

\noi However,

$$\ell n \left ( 2 + (2n+1) \exp {1 \over g_n} \right ) < {\ell n3
\over \ell n2} \left [ \ell n (2n + 1)
+ {1 \over g_n} \right ] \ ,$$

\noi and hence
$$\sum_{n=1}^{\infty} \int |V_n| \ \left | \ell n (2 +
|\vec{x}|)\right |^{\alpha}	\ d^2x  < 2 \pi \left (
{\ell n3 \over \ell n2}\right )^{\alpha} \left [ \Sigma g_n\left
[\ell n (2n + 1) + {1 \over g_n} \right
]^{\alpha} \right ] \ .$$

\noi Since we want the series on the right-hand side to converge,  
$\alpha$ is chosen to be less than $1$.
\par

With the choice

$$g_n = g_0 \exp (- \lambda n) \ ,$$

\noi this series will converge for any $\alpha < 1$. \\

\newpage
\noi {\large\bf III. Bounds on the number of bound states in one and two}\par
{\large \bf \quad dimensions} \\

We start by considering the one dimensional case, and write always,
in obvious notations, $V = V^+-V^-$ where
$V^+$ and $V^-$ are both $\geq 0$. \par

The zero-energy one-dimensional Schr\"odinger equation is

\beq
			\label{23e}
\left ( - {d^2 \over dx^2} + V (x) \right ) \psi (x)  = 0\ , \quad
x\in (- \infty , + \infty ) \ .
\eeq

Except for the fact that one is restricted to the half line, the
above equation is the same as the
reduced $\ell = 0$, 3-dimensional Schr\"odinger equation

\beq
			\label{24e}
\left ( - {d^2 \over dr^2} + V (r) \right ) u(r)  = 0\ , \quad r\in
[0 , \infty ) \ .
\eeq

Now if, in the one dimensional case $V(x)$ has $N$ bound states, then
$\psi (x)$ has $N$ nodes, $x_p$, $p
= 1, \cdots , N$. Let $k$ be such that

$$x_k < 0 < x_{k+1} \ ,$$

\noi then the 3-dimensional potential, $V_1(r) = V(x)$ with $r \equiv
x - x_{k+1}$, has $(N - k - 1)$
$\ell = 0$ bound states. Also the potential, $V_2(r) \equiv V(x)$
with $r = - (x - x_k)$ has $k$ bound
states with $\ell = 0$. Hence any three dimensional bound gives a one
dimensional bound. \par

Starting with the well known Bargmann {\cite{21r} bound for angular  
momentum $\ell$, we write

\beq
			\label{25e}
N(\ell ) < {1 \over 2 \ell + 1} \int_0^{\infty} r \, V^-(r)\, dr \ .
\eeq

Using $\ell = 0$, we get, for the one dimensional case~:
$$N(1D) - 1 < \int_{-\infty}^{x_{\kappa}} |x - x_{\kappa}| \,
V^-(x)\, dx + \int^{\infty}_{x_{\kappa + 1}} |x -
x_{\kappa + 1}|\, V^-(x)\, dx \ ,$$

\noi and hence

\beq
			\label{26e}
N(1D) < 1 + \int_{-\infty}^{+\infty} |x| \, V^-(x)\, dx \ .
\eeq

\noi Similarly, we can use the bound obtained by one of us \cite{22r}
in the radial three-dimensional
case~:

  \beq
			\label{27e}
N(3D, \ell = 0) < \left [ \int_0^{\infty} r^2 V^-(r)\, dr
\int_0^{\infty} V^-(r)\, dr \right ]^{1/4}
\eeq

\noi to get, in the one-dimensional case, after some manipulations:

  \beq
			\label{28e}
N(1D) < 1 + \sqrt{2} \left [ \int_{-\infty}^{+\infty} x^2 \, V^-(x)\,
dx \int_{-\infty}^{+\infty} V^-(x)\, dx
\right ]^{1/4} \eeq

\noi which behaves like $\sqrt{g}$ if $V = gV$, like the
semi-classical estimate.\par

Now to get bounds in two dimensions for the $m=0$ case is very
simple. The change of variables given in
Eq. (\ref{13e}) allows us to go from Eq. (\ref{26e}) to a bound for  
the 2D case~:

\beq
			\label{29e}
N(2D, m=0) < 1 + \int_0^{\infty} r  | \ell n \left ( {r \over R}\right ) | 
\  V^- (r)\,
dr \ .
\eeq

\noi In this bound $R$ is arbitrary. We can minimize with respect to
$R$. $R_{min}$ is given by

  \beq
			\label{30e}
\int_0^{R_{min}} x |V(x)|dx = \int^{\infty}_{R_{min}} x |V(x)|dx \ .
\eeq

The bound (\ref{29e}) with $R = R_{min}$ should be compared with the
bound previously obtained by Newton
\cite{13r} and Set\^o \cite{14r} which is

\bea
			\label{31e}
N(m=0) &<& 1 + {{1\over 2} \int r \ dr \ r' \ dr' V^-(r) V^-(r')
|\ell n \left ( {r \over r'}\right ) |
\over \int r \ dr \ V^-(r)} \nn \\
&=& 1 + J \ . \eea

\noi It turns out that

\beq
			\label{32e}
J < I(R_{min}) < 2J \ .
\eeq

This is demonstrated in Appendix III. So the Newton-Set\^o bound is
slightly better but has a more complex
structure. Both bounds are ``optimal'' in the sense that multiplying
factors in them cannot be improved.
This is because the Bargmann bound is itself known to be optimal.\par

Applying the same change of variable in equation (\ref{13e}) and
(\ref{28e}) gives

\beq
\label{33e}
N(m=0, 2D) < 1 + \sqrt{2} \left [ \int_0^{\infty} (\ell n r)^2 r \ dr
\, V^-(r) \int_0^{\infty} r\ dr
\, V(r)\right ] \ .
\eeq

\noi For large coupling this behaves like $\sqrt{g}$ for a potential
$gV$. The integrals appearing in Eq.
(\ref{33e}) are those which were required to converge in our original
paper on low energy scattering in 2
dimensions. \\

\noi {\large\bf IV. A bound on the total number of bound states in
two dimensions} \\

In this section, we study the total number of bound states in two  
dimensions, mostly for a rotationally
symmetrical potential. The bound for this rotationally symmetrical  
case gives also some information for
the general case, as discussed near the end of this section.

For the radial case, the easiest thing to do is to
notice that the radial reduced equation
(\ref{11e}) can be viewed as a radial three-dimensional equation with
non-integer angular momentum $\ell
= m - 1/2$. Therefore the Bargmann bound \cite{18r} is valid~:

\beq
			\label{34e}
N_m < {1 \over 2m} \int_0^{\infty} r\, V^-(r)\, dr \ .
\eeq

\noi To get the total number of bound states, we must remember that
for $m \not= 0$ we have a
multiplicity 2 and for $m = 0$ multiplicity 1. Hence

\bea
			\label{35e}
N_{total} &<& N_0 + \sum_{m=1}^{m=2\int rV^-(r)dr} {1 \over m} \int
r\, V^-(r)\, dr \nn \\
N_{total} &<& N_0 + \left [ \int r \, V^-(r)\, dr\right ] \ell n
\left [ 2 + 2 \int r\, V^-(r)\, dr\right ]
\eea

\noi where $N_0$ is for instance given by (\ref{29e}). \par

However, the logarithm is spurious. This has already happened in the
past, for instance in the
three-dimensional bound obtained by Glaser, Grosse, Martin and
Thirring \cite{21r}. \par

To show this, we use a technique due to Glaser, Grosse and Martin
\cite{9r}, in which the counting of
bound states for a radial potential is reduced to the calculation of
a bound on the moment of the
eigenvalues of a one-dimensional problem. \par

The reduced radial Schr\"odinger equation for bound states

\beq
			\label{36e}
\left [ - {d^2 \over dr^2} + {m^2 - 1/4 \over r^2} + V(r) - E_i(m)
\right ] u_i(r) = 0 \ ,
\eeq

\noi where $i$ designates the number of nodes of the solution ($i$-th
eigenfunction starting from the ground
state designated by $i = 0$), has been generalized by Regge \cite{23r}  
to non-integer and
even complex angular momentum. What can be
shown, under the weak condition

\beq
			\label{37enew}
\int r |V(r)| dr < \infty \ ,
\eeq

\noi is that each $E_i(m)$, $i = 0, 1, \cdots$ is the restriction to
$m$ integer (physical) of a real analytic,
monotonically increasing function of $m$, $0 < m < m_i$, where $m_i$
is such that $E_i(m_i) = 0$. That $m_i$
exist follows from the Bargmann bound and condition (\ref{37enew}).
(Notice that $m_0 > m_1 > \cdots$). This
is what is called a ``Regge trajectory''. Different trajectories with
different $m_i$'s do not intersect, due to
general Sturn-Liouville theory. In Appendix IV, we sketch the proof
of these statements. \par

The number of bound states on a given trajectory, with $m \geq 1$,
will be $[m_i]$, where $[x]$ is the integer
part of $x$. Each of those bound states with $m \not= 0$ has a
multiplicity~2. So the total number of bound
states with $m \not= 0$ is

$$2 \sum_{i,[m_i]\geq 1} [m_i] \ .$$

\noi On the other hand, by using the change of variables (13) already  
employed
in sections II and III
the zero-energy reduced Schr\"odinger equation
\beq
\left ( - \frac{d^2}{dr^2} + \frac{m^2 - 1/4}{r^2} + V (r) \right )
\; u(r) = 0
			\label{38e}
\eeq
\noi becomes

\beq
			\label{39e}
\left ( - {d^2 \over dz^2} + U(x)\right ) \phi (x) = - (m^2 - 1/4 ) \phi (x)
\eeq

The eigenvalues of (39) are just the $m^2_i - 1/4$, $m_i$ defined
previously.
\noi The sum $\sum[m_i]$ is very similar to the sum of moments of
power 1/2 of the eigenvalues of
(38) :

\beq
			\label{40e}
\sum_{[m_i] > 1} [m_i] < {2 \over \sqrt{3}} \sum (m_i^2 - 1/4)^{1/2} \ .
\eeq

\noi It happens that this moment satisfies a bound proposed by Lieb
and Thirring \cite{24r}

\beq
			\label{41e}
\sum |e_i|^{1/2} < L_{1/2,1} \int_{-\infty}^{+\infty} dx\, U^-(x) =
L_{1/2,1} \int_0^{\infty} r\, V^-(r)\, dr
\eeq.
where the $e_i$'s are the eigenvalues of the one-dimensional  
Schr\"odinger equation with a potential $U$.
$L_{1/2,1}$ has been shown to be finite by Weidel \cite{25r} and
less than 1.005. More recently
Hundertmark, Lieb and Thomas \cite{26r} have found the optimal value
for $ L_{1/2,1}$, namely 1/2~:

\beq
\label{42e}
\sum |e_i|^{1/2} < {1 \over 2} \int_{-\infty}^{+\infty} U^-(x)\, dx
\eeq

\noi which is obtained in the one-bound-state case with a delta
function potential. \par

Therefore, using (\ref{29e}), (\ref{40e}) and (\ref{42e}) we get a
bound on the total number of bound
states in two space dimensions for a central potential

	 \bea
			\label{43e}
N < 1 &+& \int_0^{\infty} r\, V^-(r) \ \left | \ell n \left ( {r
\over R}\right ) \right |dr\nn \\
&+& {2 \over \sqrt{3}} \int_0^{\infty} r\, V^-(r)\, dr \ .
\eea

\noi We notice that for a potential $gV$ the bound is {\it linear} in
$g$, similar to the semiclassical
estimate for large $g$. It is probably almost optimal, in the sense
that it is optimal for $m = 0$
and that for $m \not= 0$ the only foreseeable improvement is to remove  
the multiplicative factor
$2/\sqrt{3}$.\par

It is trivial, but not very elegant, to obtain also a bound on the
total number of bound states for a
non-central potential. Let

	 \beq
			\label{44e}
B(r) = \sup_{0 < \theta < 2 \pi} \ V^-(r, \theta ) \ .
\eeq

\noi Then replacing $V(r)$ by $B(r)$ in (\ref{43e}) we get a bound on
the total number of bound
states in a non-radial potential because of the monotonicity of the
bound-state energies with respect
to the potential.\par

For a potential with a single singular point the replacement of $V^-$
by $B(r)$ is not too bad.
However, if $V$ has several singular points the replacement will be
catastrophic since $B$ will be
infinite on successive circles corresponding to these singular
points. It is certainly desirable to
find a better bound. \par

Our conjecture is

\bea
			\label{45e}
N &<& 1 + 2 \int {d^2x \over 2 \pi} \ V_R(|x|) \ell n^{-} \left ( {|x| 
\over R} \right )\nn \\
&+&  \int {d^2x \over 2 \pi} \ V^-(x) \, \ell n \left ( {|x| \over R}
\right ) + {2 \over \sqrt{3}} \int \frac{d^2x}{2 \pi} \
V^-(x)\ , \eea

\noi where $V_R(|x|)$ is the decreasing rearrangement of $V^-(x)$
(see footnote in Appendix I). The reasons for
which we propose this are \par

(i) for a central decreasing potential (\ref{45e}) coincides with
(\ref{43e})~;\par

ii) for a central potential not necessarily decreasing, the r.h.s. of
(\ref{45e}) is larger than the r.h.s. of
(\ref{43e})~; \par

iii) if we take a shifted central with a centre \underbar{outside the
origin}, the first and the last integrals
in (\ref{45e}) are, of course, invariant. The second integral,
because of the harmonic properties of $\ell n\,
r$ in 2 dimensions, is larger than the one corresponding to a central
potential centred at the origin. \par

Proving (\ref{45e}) or something similar might be rather difficult
but, seeing what has been achieved
for higher dimensions, not impossible. \par

Notice that the integrals in (\ref{43e}) and (\ref{45e}) will
certainly converge under the conditions
of Ref. \cite{1r}, and we can announce that they do converge in Ref.
\cite{2r} also. \\

\noi {\large \bf Acknowledgements} \par

We should like to thank J. M. Richard and W. Thirring for crucial  
information. Two of us, N.N.K and T.T.W.,
are grateful to the CERN Theory Division for its kind hospitality.  
This work was supported in part by the U.S. Department
of Energy under Grant No. DE-FG02-91ER40651, Task B, and under Grant  
No. DE-FG02-84ER40158.\\
\\
\noi {\it Note added:} Dr. P. Blanchard drew our attention to a paper  
by A. Laptev \cite{28r} in which he finds a bound
on the number of bound states for a potential
$b|x|^{-2} - |V(|x|)|$, which is
$$
N < \frac{A(b)}{4 \pi} \; \int|V(x)|d^2 x ,
$$
when $A(b) \to \infty$ for $b \to 0$. With methods developed in the  
present paper, using the Bargmann bound for the $m =
0$ contribution and (42) for the rest, we get
$$
A(b) < \frac{1}{\sqrt{b}} +  \frac{4}{\sqrt{3}}.
$$

\newpage

  \begin{center}{\large \bf Appendix I}\vskip 3 truemm

{\large \bf  Upper and lower bounds on the ground state energy}\\
{\large \bf in two
dimensions}
\end{center}
\vskip 1 truecm

We use the Schr\"odinger equation in integral form, for a potential $gV$~:

$$\psi (x) = -{g \over 2 \pi} \int K_0 (\kappa |x - y|) V(y) \psi (y)
d^2y \ , \eqno({\rm I}.1)$$

\noi for an energy $E = - \kappa^2$. \par

First we shall get an algebraic lower bound. Then $V$ can be replaced
by $- V^{-}$, the attractive
part of the potential. We have~:

$$|\psi (x) | < {g \over 2 \pi} \int K_0 (\kappa |x - y|) V^-(y) d^2y
\sup |\psi | \ . \eqno({\rm
I}.2)$$

\noi Since $K_0(t)$ is a decreasing function of $t$ and given the
rearrangement inequality,

$$\int AB \ d^2x < \int A_R\ B_R \ d^2x $$

\noi where $A$ and $B$ are positive, going to zero at infinity, and
$A_R$ and $B_R$ are their
decreasing circular rearrangements\footnote{$A_R$ is a decreasing
function of $|x|$ such that $\forall
\ t$, $\mu (A_R > t) = \mu (A > t)$, where $\mu$ is the Lesb\`egue
measure.}, we have

$$|\psi (x) | < {g \over 2 \pi} \int K_0 (\kappa |y|) V_R(|y|) d^2 y
\sup |\psi | \eqno({\rm I}.3)$$

\noi where $V_R$ is the rearrangement of $V^-$. Hence, if we take the
supremum of the left-hand side
over $x$, we can divide by $\sup |\psi |$ and obtain

$$1 < {g \over 2 \pi} \int K_0 (\kappa |y|)V_R(|y|)d^2y \ .$$

\noi From the property

$$K_0 (ab) < \ell n^+ \left ( {1 \over a}\right ) + K_0 (b) \
,\eqno({\rm I}.4) $$

\noi where $\ell n^+(t) = \ell n \ t$ for $t > 1$, $= 0$ for $t < 1$,
which is proved at the end of
this Appendix, we get

$$K_0(\kappa ) > {1 \over g} \ {1 - {g \over 2 \pi} \int \ell n^+
\left ( {1 \over |y|}\right ) V_R(y)
d^2 y \over {1 \over 2 \pi} \int V_-(y) d^2y} = X \ . \eqno({\rm I}.5)$$

\noi As long as $X$ is positive, this gives a lower bound on
$K_0(\kappa )$ and hence an upper bound
on $\kappa$ and an upper bound on $\kappa^2$, the absolute value of
the binding energy. \par

If $X > K_0 (1) = 0.42, \cdots$, we can again use the inequality (I.4)  
and get

$$\kappa^2 < \exp 2 \left ( - {1 \over g} \ {1 - {1 \over 2 \pi} \int
\ell n^+ \left ( {1 \over
y}\right ) V_R(y) d^2 y \over {1 \over 2 \pi} \int V_-(y) d^2y} +
K_0(1) \right ) \ , \eqno({\rm I}.6)$$

\noi which demonstrates that the absolute value of the binding energy
is bounded by $\exp - C/g$, $C >
0$ for $g \to 0$. \par

In the special case of a {\it purely attractive} potential we can get
an inequality going in the
opposite direction. We start again from (I.1) and use the fact that
the ground-state wave function is
positive. We have

$$\psi (x) > {g \over 2 \pi} \int_{|y|< R} K_0(\kappa |x-y|) |V(y)
|d^2y \times {\rm Inf}_{|y| < R} \ |\psi
(y)|$$

\noi and, taking also $|x| < R$, and using the fact that $K_0$ is  
decreasing~:

$${\rm Inf} |\psi (y)|_{|x| < R} > {g \over 2 \pi}\ K_0(2\kappa R)
\int_{|y|< R} |V(y) |d^2y \ {\rm Inf} |\psi
(y)|_{|x| < R} \ . \eqno({\rm I}.7)$$

\noi However, ${\rm Inf} |\psi (y)|_{|x| < R}$ cannot vanish in the
ground state and hence we can
divide (I.7) by ${\rm Inf} |\psi (x)|$. From

$$K_0(t) > \ell n \ {1 \over t} + \ell n2 - \gamma \ , \eqno({\rm I}.8)$$

\noi when $\gamma$ is the Euler constant $= 0.577$ ... we get

$$\kappa^2 > {e^{-2\gamma} \over R^2} \ \exp - {2 \over g \int_{|x|<
R} |V(\kappa )|d^2x}
\eqno({\rm I}.9)$$

\noi which goes in the opposite direction to (I.6), but again has the
form $\exp - {C \over g}$ for
small $g$. Both upper and lower bounds on $\kappa^2$ have the same
qualitative behaviour for small
$g$. The lower bound on $\kappa^2$ can be optimized with respect to
$R$. Of course we cannot do that
for a potential which is not strictly attractive but only globally
attractive. Nevertheless, we
believe that the same qualitative result will hold.\par

In a recent paper \cite{27r} Nieto has given an explicit example in
which he shows that the binding
energy in absolute value is incredibly small. A square well with
unit radius and strength 0.1 in
natural units produces a bound state with energy $- 10^{-18}$. \par

Finally we give a proof of (I.4) and (I.8)~: consider the quantity

\begin{eqnarray*}
&&Z = K_0(x) - \ell n \left ( {x_0 \over x}\right ) \ , \\
&&Z' = - K_1(x) + {1 \over x} \ .
\end{eqnarray*}

\noi From

$$K_1(x) = \int_1^{\infty} {t dt \over \sqrt{t^2-1}} \ \exp - tx <
\int_1^{\infty} {tdt \over
\sqrt{t^2-1}} \ \exp - x \sqrt{t^2 - 1} \ , $$

\noi we get $K_1(x) < {1 \over x}$, and hence

$$ Z' > 0 \ .$$

\noi So, for $x < x_0$ $Z(x) < Z(x_0) = K_0 (x_0)$, which proves
(I.4). On the other hand, we have
$\lim\limits_{x \to 0} Z(x) = \ell n2 - \gamma$, and so

$$K_0(x) > \ell n2 - \gamma + \ell n \left ( {1 \over x} \right ) \ .$$

\newpage

  \begin{center}{\large \bf Appendix II}\vskip 3 truemm

{\large \bf  Transformations of the Schr\"odinger equation}\\
{\large \bf   from one to two dimensions,}\\ {\large \bf
the converse, limit potentials, and generalization} \end{center}
\vskip 1 truecm

In section II we presented a transformation of the one dimensional
zero energy Schr\"odinger equation,

$$\left ( - {d^2 \over dx^2} + U(x) \right ) \phi ( \kappa ) = 0 \ ,
\quad x \in (- \infty , + \infty
) \ ; \eqno({\rm II}.1)$$

\noi into the two dimensional, zero angular momentum, Schr\"odinger equation,

$$\left ( - {d^2 \over dr^2} - {1 \over r } \ {d \over dr} + V(r)
\right ) \phi ( r ) = 0 \ , \quad
r \in [0, \infty ) \ . \eqno({\rm II}.2)$$

The transformation is given by~:

\begin{eqnarray}
\hspace*{2cm} x &\equiv & \ell n \left ( {r \over R} \right ) \ ;  
\quad x \in (-
\infty , + \infty ) \ ; \quad r \in
[0, \infty ) \ ; \nonumber \\
\hspace*{2cm} U(x) &\equiv & r^2\ V(r) \ ; \qquad \ x \geq 0 \ ; \nonumber \\
\hspace*{2cm} \phi (x) &\equiv & \psi (r) \ ; \qquad \qquad x \geq 0 \  
. \nonumber
\hspace*{7cm}({\rm II}.3)
\end{eqnarray}

This enables us to prove that since a potential, $U(x)$, given by

\begin{eqnarray}
\hspace*{2cm} U(x) &=& 0 \ ;  \quad x < X \ ,\nonumber \\
\hspace*{2cm}
U(x) &=& - {\mu \over 4x^2} \ ; \quad \mu > 1\ , \ x \geq X \ , \nonumber  
\hspace*{6.4cm}({\rm II}.4)
\end{eqnarray}

\noi has infinitely many bound states in one dimension, the potential

\begin{eqnarray}
\hspace*{2cm}  V(r) &=& 0 \ ; \qquad \quad r < R_0 \ ;\nonumber \cr
\hspace*{2cm} V(r) &=& - {\mu \over r^2\left (\ell n {r\over R} \right  
)^2} \ ;
\quad r \geq R_0 > R \ ; \ \mu > 1 \ ;
 \hspace*{4.5cm} ({\rm II}.5) \nonumber
\end{eqnarray}

\noi will also have infinitely many bound states in two dimensions
for the $m = 0$, radial case. \par

This procedure can be continued further. We can re-transform (II.1)
to make it look like a two
dimensional equation by defining $\chi (x)$ as

$$\phi (x) \equiv x^{1/2} \ \chi (x) \ . \eqno({\rm II}.6)$$

\noi The $\kappa$ satisfies the equation

$$\left ( - {d^2 \over dx^2} - {1 \over x } \ {d \over dx} + W(x)
\right ) \chi ( x ) = 0 \ ,$$

\noi with

$$W(x) = U(x) - {1 \over 4x^2} \ . \eqno({\rm II}.7)$$

\noi Relabelling $x$ as $r$ we have

$$\left ( - {d^2 \over dr^2} - {1 \over r } \ {d \over dr} + W(r)
\right ) \chi ( r ) = 0 \ . \eqno({\rm
II}.8)$$

\noi This last equation is for $r \geq 0$ exactly the two dimensional
radial equation. \par

 From the chain,

$$V(r) \to U(x) \to W(r) \ ,$$

\noi we obtain,

$$W(r) = - {1 \over 4r^2 (\ell n \ r)^2} + {1 \over r^2} \ V(\ell n \
r ) \ . \eqno({\rm II}.9)$$

\noi Thus if for $x > x_0$ we set

$$U(x) = - {\mu \over 4x^2} \ ,$$

\noi or

$$V(r) = - {\mu \over 4r^2 (\ell n\ r)^2} \ ,$$

\noi we get

$$W(r) = - {1 \over 4r^2 (\ell n\ r)^2} - {\mu \over 4r^2 (\ell n \
r)^2(\ell n \ \ell n \ r)^2} \ ,$$

\noi with $r > R_0 > 0$. \par

This potential has infinitely many bound states if $\mu > 1$. Our
procedure can be repeatedly iterated
producing potentials which are closer to the limit, and with wave
functions which can be expressed
explicitly in terms of elementary functions. \par

Finally we stress that this procedure is not restricted to the
connection between one dimension and
two dimensions, and the construction of limit potentials in one or
two dimensions. It also applies in
$N$ dimensions. \par

In $N$-dimensions the radial Schr\"odinger equation becomes

$$\left ( - {d^2 \over dr^2} - {N - 1 \over r } \ {d \over dr} + V(r)
\right ) \psi ( r ) = 0 \ .$$

\noi We set

$$\psi (r) = r^{1 - {N \over 2}} \ \widetilde{\psi}(r) \ ;$$

\noi and obtain

$$\left ( - {d^2 \over dr^2} - {1 \over r } \ {d \over dr} + {(1 - {N
\over 2})^2 \over r^2} + V(r)
\right ) \widetilde{\psi} ( r ) = 0 \ .$$

\noi We can define $\widetilde{V}(r) \equiv V(r) + \left ( 1 - {N
\over 2} \right )^2/r^2$, and hence
again obtain the 2-$D$ form. \par

The conclusion is, using (II.5), that in $N$ dimensions, the potential

\begin{eqnarray}
\hspace*{3cm}  V(r) &=& - {(N-2)^2 \over 4r^2} - {\mu \over r^2(\ell n  
\ r)^2}  \ ;
\qquad  r > R > 1\ ;\nonumber \cr
\hspace*{3cm} &=& 0 \quad ; \qquad r \leq R \ ;  \nonumber
\hspace*{7cm} ({\rm II}.11)
\end{eqnarray}

\noi has infinitely many bound states if $\mu > 1$, and a finite
number if $\mu < 1$. \par

This procedure can be further iterated to get more refined results.

\newpage

  \begin{center}{\large \bf Appendix III}\vskip 3 truemm
  {\large \bf  Comparison of our bound on the number}\\
{\large \bf of m = 0 bound states and of the Newton-Set\^o bound}
\end{center}
\vskip 1 truecm

We wish to compare our bound

\begin{eqnarray}
\hspace*{3cm}  N(m = 0) &<& 1 + I(R) \;  \nonumber \cr
\hspace*{3cm} I(R) &=& \int_0^{\infty} r dr V^- (r) \left |\ell n  
\left ( {R \over
r} \right ) \right | \, \nonumber
\hspace*{5cm} ({\rm III}.1)
\end{eqnarray}

\noi and $I(R_{min})$ given by

$$\int_0^{R_{min}} r \ V^-(r) dr = \int_{R_{min}}^{\infty} r \ V^-
(r) dr \ ,\eqno({\rm III}.2)$$

\noi with the Newton-Set\^o bound~:

$$N(m = 0) < 1 + J ,$$

\noi where

$$J = {{1 \over 2} \int \int r\ dr\ r'\ dr' \left | \ell n \left ( {r
\over r'} \right ) \right
| V^-(r)\ V^-(r') \over \int r \ dr \ V^-(r)}. \eqno({\rm III}.3) $$

\noi $J$ can be rewritten as

$$J = {{1 \over 2} \int r\ dr\  V^-(r)\ I^-(r) \over \int r \ dr \
V^-(r)} \ . \eqno({\rm III}.4)$$

\noi Hence, from the mean value theorem

$$J \geq {1 \over 2} \ I(R_{min}) \ . \eqno({\rm III}.5)$$

\noi On the other hand, taking into account (III.2), one has, with
$R > R_{min}$,

$$I(R) = I(R_{min}) + 2 \int_{R_{min}}^R r\ dr\ V^-(r) \ \ell n \left
( {R \over r} \right ) \ . \eqno({\rm
III}.6)$$

\noi One gets

\begin{eqnarray}
I(R) &<& I(R_{min}) + 2 \ell n \left ( {R \over R_{min}}\right )
\int_{R_{min}}^{\infty} r\ dr\ V^-(r)
\nonumber \cr
&=& I(R_{min}) + \ell n \left ( {R \over R_{min}}\right )
\int_0^{\infty} r\ dr\ V^-(r) \ . \nonumber
\end{eqnarray}

\noi The case $R < R_{min}$ can be treated in the same way and one gets

$$I(R) < I(R_{min}) + \left |  \ell n \left ( {R \over R_{min}}\right
) \right | \int_0^{\infty} r\
dr\ V^-(r)  \ . \eqno({\rm III}.7)$$

\noi Inserting in (III.4) leads to

$$J < I(R_{min}) \ . \eqno({\rm III}.8)$$

\newpage

  \begin{center}{\large \bf Appendix IV}\vskip 3 truemm
{\large \bf  Regge trajectories for bound states}\\
  \end{center}
\vskip 1 truecm

What follows here is somewhat implicit in the work of Regge
\cite{23r}. We give here some details for the sake
of completeness. \par

To find bound state energies $E = - \kappa^2$ for a given $m$ (real $> 
0$), but not necessarily integer, we must
find a solution of

$$\left [ - {d^2 \over dr^2} + {m^2 - 1/4 \over r^2} + V(r) + \kappa^2 
\right ] u = 0 \ ,\eqno({\rm IV}.1)$$

\noi such that $u \to 0$ for $r\to 0$ and $r \to \infty$. For general
$m$ and $\kappa$ ,\, $Re\, m > 0$, \, $Re\ \kappa >
0$, if

$$\int r |V(r)|dr < \infty \ ,\eqno({\rm IV}.2)$$

\noi (IV.1) has in general two independent solutions $y$ and $z$ such that

\begin{eqnarray}
\hspace*{3cm}  y &\sim & r^m \qquad , \qquad r \to 0 \qquad \nonumber \cr
\hspace{3cm}  z &\sim & \exp (-\kappa r) \qquad , \qquad r \to \infty \ .
 \, \hspace*{5.5cm} ({\rm IV}.3) \nonumber
\end{eqnarray}

\noi It is then shown that both $y(m, \kappa ; r)$ and $z(m, \kappa ; r)$ 
are analytic in $m$ and $\kappa$ in \\
$\{Re\ m
> 0 \otimes Re\ \kappa > 0\}$. The Wronskian of $y$ and $z$ is given by

$$W(y, z) \equiv yz' - y'z = F(m, \kappa ) \ ,$$

\noi where $F$ is analytic in the same domain. The bound state
energies are given by

$$F(m, \kappa ) = 0 \ .\eqno({\rm IV}.4)$$

\noi This defines the bound state energies as implicit functions of
$m$. If $F(\widetilde{m}_i,
\widetilde{\kappa}_i) = 0$, $\widetilde{m}_i$ and
$\widetilde{\kappa}_i >0$, and $(\partial / \partial \kappa )^p F
= 0$, $p = 1, 2 \cdots , q-1$, and $(\partial / \partial \kappa )^q F
\not= 0$ at that point, we have $q$ different solutions in the
neighbourhood of $\widetilde{m}_i$,
$\widetilde{\kappa}_i$. However, this is impossible for $q \geq 2$
because there cannot be any degeneracy as a
general consequence of Sturm-Liouville theory. Hence, $\kappa$ is
analytic in $m$ in the neighbourhood of $\widetilde{m}_i$,
$\widetilde{\kappa}_i$, and $\kappa_i$ is a real analytic function of $m$ 
for $0 < m < m_i$, where $m_i$ is such
that $E_i (m_i) = 0$. In addition, $\kappa_i$ is a decreasing function 
of $m$ since, from the Feynman-Hellmann
theorem

$${dE_i \over dm} = 2m \int {u^2_i \over r^2}\ dr  \ .\eqno({\rm IV}.5)$$

\noi Let us remark here that the condition (IV.2) is certainly too
strong. It is needed to ensure that $y$ and
$z$ have the properties given by (IV.3). But if $V$ has strong
repulsive singularities, one could approach it by
$V_M, V_M = V$ if $V
< M$, $V_M = M$ if $V \geq M$, and use a
limiting procedure.


\newpage
\begin{thebibliography}{99}
\bibitem{1r} F. Bassani, T. Martin, private communications.

\bibitem{2r} K. Chadan, N. N. Khuri, A. Martin and T. T. Wu, {\it
Phys. Rev.} {\bf D58} (1998) 025014.

\bibitem{3r} N. N. Khuri, A. Martin, P. Sabatier and T. T. Wu, in  
preparation.

\bibitem{4r} K. Chadan, {\it Nuovo Cimento} {\bf 58A} (1968) 191.

\bibitem{5r} A. Martin, {\it Helv. Phys. Acta} {\bf 45} (1972) 140.\\
H. Tamura, {\it Proc. Japan. Acad.} {\bf 50}
(1974) 19.

  \bibitem{6r} E. Lieb, {\it Bull. Amer. Math. Soc.} {\bf 82} (1978)
751 and {\it Proc. A.M.S., Symp.
Pure Math.} {\bf 36} (1980) 241.

\bibitem{7r} M. Cwikel, {\it Trans. AMS} {\bf 224} (1977) 93.

\bibitem{8r} G. V. Rozenbljum, {\it Dokl. AN SSSR} {\bf 202 NS}
(1972) 1012~; Izv. {\it VUZOV
Mathematika} {\bf N1} (1978) 75.

\bibitem{9r} H. Grosse, V. Glaser and A. Martin, {\it Comm. Math.
Phys.} {\bf 59} (1978) 197.

  \bibitem{10r} P. Li and S. T. Yau, {\it Comm. Math. Phys.} {\bf 88}
(1983) 309~; \\
Ph. Blanchard, J. Stubbe and J. Rezende, {\it Lett. Math. Phys.} {\bf
14} (1987) 215~;\\
J. G. Conlon, {\it Rocky Mountain J. Math.} {\bf 15} (1985) 117.


\bibitem{11r} B. Simon, {\it Ann. Phys.} {\bf 97} (1976) 279.


\bibitem{12r} K. Yang and M. de Llano, {\it Am. J. Phys.} {\bf 57} (1989) 85.


\bibitem{13r} R. G. Newton, {\it J. Math. Phys.} {\bf 3} (1962) 867.

\bibitem{14r} N. Set\^o, {\it Publ. RIMS}, Kyoto University (1979) 429.

\bibitem{15r} T. Regge, {\it Nuovo Cimento} {\bf 14} (1959) 951.

\bibitem{16r} N. N. Khuri, A. Martin and T.T. Wu, {\it Few-Body  
Systems} {\bf 31} (2002) 83-89.

\bibitem{17r} R. Courant and D. Hilbert, Methods of Mathematical
Physics, vol. I, Interscience, New York (1953)
p. 454.

\bibitem{18r} F. Calogero, {\it Comm. Math. Phys.} {\bf 1} (1965) 80.\\
H. E. Cohn, {\it J. London Math. Soc.} {\bf 40} (1965) 523.

\bibitem{19r} See for instance, E. Hille, Lectures on Ordinary
Differential Equations, Addison-Wesley (1969), p.
340.

\bibitem{20r} J. M. Richard, private communication.

\bibitem{21r} V. Bargmann, {\it Proc. Nat. Acad. Sci.} USA {\bf 38}  
(1952) 96.

\bibitem{22r} A. Martin, {\it Comm. Math. Phys.} {\bf 55} (1977) 293.

\bibitem{23r} V. Glaser, H. Grosse, A. Martin and W. Thirring, in
``Studies in Mathematical Physics'', essays in
honour of V. Bargmann, E. Lieb, B. Simon, A. Wightman eds., Princeton
University Press, Princeton (1978), p. 169.

\bibitem{24r} E. H. Lieb and W. Thirring, Ref. [22], p. 269.

\bibitem{25r} T. Weidel, {\it Comm. Math. Phys.} {\bf 178} (1996) 135.

\bibitem{26r} D. Hundertmark, E. H. Lieb and L. E. Thomas, {\it Adv.
Theor. Phys.} {\bf 2} (1998) 719.

\bibitem{27r} M. Nieto, {\it Phys. Letters} {\bf 293} (2002) 10.

\bibitem{28r} A. Laptev, {\it Functional Analysis and its  
Applications} {\bf 34} (2000) 305.
\end{thebibliography}
\end{document}